# Assembling biological boolean networks using manually curated databases and prediction algorithms

Alberto Calderone

*Institute for Systems Analysis and Computer Science "Antonio Ruberti", Rome, Italy*
email: sinnefa@gmail.com

Despite the large quantity of information available, thorough researches in various biological databases are still needed in order to reconstruct and understand the steps that lead to known or new phenomena. By using protein-protein interaction networks and algorithms to extract relevant interconnections among proteins of interest, it is possible to assemble sub-networks from global interactomes. Using these extracted networks it is possible to use algorithms to predict signal directions while activation and inhibition effects can be predicted using RNA interference screenings. The result of this approach is the automatic generation of boolean networks. This way of modelling dynamical systems allows the discovery of steady states and the prediction of stimuli response.

## INTRODUCTION

Systems biology is a discipline that studies natural phenomena as complex systems trying to answer to biological questions exploiting the system at study in its entirety. This approach has been very useful to answer many questions about the dynamics and the organisation of living thing. Recently, the ever-growing amount of data available combined with computational approaches allowed the predictions and analysis of complex models, in the most diverse situations.

One of the main challenges of systems biology is the study of biological networks in order to understand the dynamics that govern their behaviours (Alon, 2007). One possible way of modelling biological systems is by creating networks of interactions in conjunction with boolean formalism where nodes can exist in two states (active and inactive) and links that activate or inactivate the target protein.

Unfortunately, signalling information currently available in databases, i.e. the direction of an interaction and the effect it has on the target protein, is sparse in different resources, limited in number and it is archived using different policies. This information needs to be merged and possibly complemented with prediction strategies in order to broaden the overall coverage of the human interactome. To this end, several computational approaches have attempted to fill the gap between protein-protein interaction networks, for which a lot of information is available, and signalling networks.

In this work, I will describe the possibility of exploiting the information contained in *mentha* (Calderone et al., 2013) – a large protein-protein interaction (PPI) database, in order to assemble signalling networks in the form of boolean graph dynamical systems. I talk about strategies to automatically add extra proteins to a starting set of interest in order to create a backbone to build signalling networks. The topology of a signed oriented network can be used to assign transition functions to each node of the network in order to use the resulting model to perform boolean dynamics simulations.

Techniques that generate high-throughput data sets are usually targeted at discovering and analysing regulatory networks (Bar-Joseph et al., 2003; Segal et al., 2003) and metabolic networks (Covert et al., 2004) in order to reconstruct signalling networks and pathways (Bebek and Yang, 2007; Scott et al., 2006). High-throughput data gives information about the directionality of an interaction but it does not give information about the cascade that occurs, for instance, from a receptor to a transcription factor and, ultimately, to the activation of one specific gene. As an example, ChIP-chip and ChIP-Seq identify which transcription factors regulate genes, or microRNAs but not the intermediate steps.

While signalling networks are directed, most of available protein-protein interaction data are undirected (Ewing et al., 2007). Even though some interactions are not oriented at all and some are known to have a specific directionality, like kinase-substrate and phosphatase-substrate, the directionality of the vast majority of protein interactions is still unknown.

Despite the fact that most of protein-protein interactions are not currently oriented, it is possible to predict this directionality using computational means. The problem of orienting interactions in a PPI network has been shown to be NP-hard (Blokh et al., 2013) and therefore several approximations have been proposed. Gitter et al. (Gitter et al., 2011) compared different approaches to computationally predict edge orientations with good results. They compared algorithms that predict signalling cascades using a set of starting point and a set or ending point in conjunction with a completely non-oriented network or a partially oriented network. This approaches are generally computationally heavy for large networks.

**STRATEGY**

Using the algorithms proposed by Gitter it is possible to exploit the sparse manually curated signalling information extracted from Kegg (Kanehisa and Goto, 2000; Kanehisa et al., 2012) and SignaLink (Fazekas et al., 2013) in order to create oriented networks. As starting points it is possible to use data from HPMR (Ben-Shlomo et al., 2003), a manually curated database of receptors and substrate and, as target points transcription factors from two manually curated databases called TcoF-DB (Schaefer et al., 2011) and AnimalTFDB (Zhang et al., 2011). Feeding these three elements to the orientation algorithms we can obtain a fully oriented network which preserves already known directionality. In particular, due to its speed and better orientations, I used the "Random+search" approach reported by Gitter et al., which tries to orient edges by randomly flipping directions in order to satisfy the highest number of path followed by a search for edges that, is flipped, improve the final result.

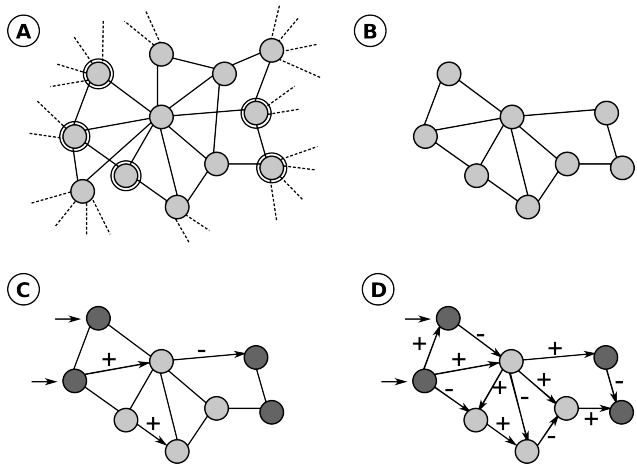

*Figure 1: Steps to generate signed oriented network. Given an interactome A we need to extract a sub-network starting from nodes of interest (circled nodes). The resulting network should be the best possible for the starting nodes. Having a sub-network B we can define starting points, for instance cellular receptors, and end points, for instance transcription factors. Secondly, from signalling databases we can extract some signs and directions and obtain the network C. Finally, applying an orientation algorithm and a sign prediction algorithm we can complete the network D and run simulations.*

Before applying an orientation algorithm, we need to produce a network which is consistent to the problem we are studying. Using a collection of proteins it is possible to extract sub-networks from a global interactome. A simple sub-network consisting of all possible neighbouring proteins may contain unnecessary proteins that would make the computation time and the complexity of the final model intractable. In general, not all interactors are important to specific signalling cascades; different isoforms or different pathways might be related to different cascades triggered by different receptors. To this end, a selective sub-network extraction approach is needed. A non exhaustive list of algorithms to extract sub-networks from a weighted interactome, such as *mentha* interactome, is the following: all-vs-all minimum path, Steiner Tree (Sadeghi and Fröhlich, 2013) and PageRank (Page et al., 1999) and emission decay (Lemetre et al., 2013).

Secondly, in order to assign to each oriented edge a sign, it is possible to use an algorithm that exploits biological clues from high throughput experiments. In particular, I used a recently proposed approach that exploits RNAi screenings (Vinayagam et al., 2014) adding, on top of the oriented network generated in the previous step, an effect (positive or negative, activation or inhibition) to each interaction. While Vinayagam approach has been evaluated on *Drosophila melanogaster* I evaluated the algorithm performances on *Homo sapiens*.

To conclude, starting from a set of protein we can extract other intermediate proteins using sub-network extraction algorithms. These nodes are liked one another through interactions that might have a direction and a sign annotated in some database. The partially oriented signed network is fed to an orientation algorithm that predicts missing edge direction. Finally, each edge will get a sign assigned with the algorithm that predicts effects using RNAi screening. Considering that proteins (nodes) can exist in two states, active and inactive, and that edges have a sign that can influence the target protein, we obtain a boolean network.

Even though the best way to validate Boolean networks would be to verify how they behave in comparison with experimental results, this approach is not applicable for automatically generated networks. In order to evaluate the oriented, signed boolean network generated by the previous steps, I implemented the recently proposed "regulation entropy" (Wu et al., 2009) which measures the number of discordant activation/inhibition signals that get to each node in a network.

The resulting model, which is signed, oriented and has valid regulation entropy, can be used as a boolean network given that each node has a transition function assigned. A possible approach, is the one that assumes the dominance of inhibitions – except for self degradation that should be overruled by activations. This approach is called "strong inhibition". An overview of the strategy is illustrated in figure 1.

**PRELIMINARY RESULTS**

In order to decide which algorithm to use to extract a sub-network from *mentha*'s interactomes, I compared three different sub-network extraction algorithms: all vs all, Steiner Tree and PageRank. All vs all extracts all possible ways of linking two protein sets while Steiner tree and PageRank extract only a fraction of them. To estimate the precision of each algorithm I used Reactome (Matthews et al., 2009) pathways as reference. The pathways I used are: EGFR pathway (265 proteins), apoptosis pathway (159 proteins) and the cell cycle pathway (496 proteins). I randomly extracted sets of proteins from these three pathways in order to perform the three algorithms separately. Each algorithm expands the starting set with other proteins (Figure 2) and, in order to decide the best algorithm, I counted the number of proteins, other than the starting proteins, that are contained in the reference pathway.

The way a pathway is defined may vary from database to database because in some cases it is not clear if a protein belongs to a pathway or not, it might depend on where one wants to draw a line between two biological processes. Consequently, I did not consider wrong the proteins that were not contained in Reactome's pathway. I calculated only *precision* and not *recall*. No matter the algorithm, the output size should be as small as possible, which always results in a small *recall*; we want the output to be small in order to ease orientation algorithms and "regulation entropy", which are both computationally heavy.

From the charts in figure 2 and 3 it can be seen that, even if the output set is smaller in Steiner, the overall precision is slightly higher, justifying the use of Steiner tree algorithm to extract sub-networks over all-vs-all minimum path algorithm. PageRank algorithm seems to have bad performances no matter the input size. Since the output of all the three algorithm contains the input proteins, I only computed precision over the extracted proteins.

To predict edges' sign I run the Vinayagam algorithm. As stated earlier, to evaluate its performances I calculated accuracy using RNAi screening for *Homo sapiens* using RNAi screenings from GenomeRNAi (http://genomernai.de/GenomeRNAi/) obtaining an *Accuracy* of 0.743, *Sensitivity* of 0.896 and *Specificity* of 0.084. The "true positive rate" (Sensitivity) is fairly good but, on the other hand the "true negative rate" (Specificity) is quite low. The overall accuracy (true positives and negatives divided by all positive and negatives) is 0.743, which is not too bad but it does introduce an amount of uncertainty in the final model generated.

Finally, one way to evaluate dynamical biological networks is regulation entropy, which is a measure of how discordant activation pathways and inhibition pathways are. An activation pathway is defined as a path whose multiplied signs (activation +, inhibition -) give as a result + (plus). On the other hand, an inhibition pathway is defined as a path whose multiplied signs give as a result - (minus). According to Wu et al. Biological networks tend to have a regulatory entropy between 0.4 and 0.7. After some tests I verified that networks generated using both manually curated data and predicted data have a better regulation entropy compared to those assembled using only manually curated data.

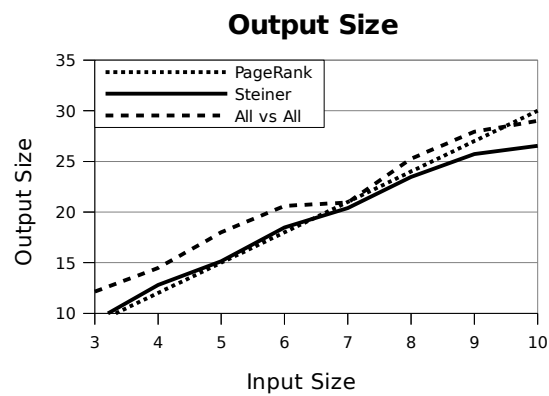

*Figure 2: Sub-network extraction output size growth rate.*

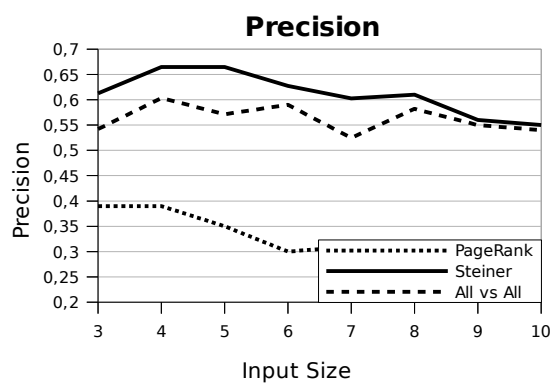

*Figure 3: Sub-network extraction algorithms precision versus input size. Decrease in precision is due to the fact that the output size increases more than the number of correctly identifies proteins.*

## CONCLUSIONS

In the short assay I wanted to show how it is possible to complement manually curated data with predicted information in order to automatically generate boolean graph dynamical systems. Using these models, it is possible to predict what proteins are active or inactive, for instance, by fixing a protein to "active", which might represent a drug response. There are still many open point like the scarce amount of RNAi screening available to predict interaction singes, or the evaluation of the best algorithm to extract interactions from PPI. Last but not least, in order to start a dynamical boolean system we still have to use approximations like the "strong inhibition" while manually curated transition rules for each node would be advisable.

Despite the aforementioned points, the results obtainable are promising. It is actually possible to set-up a pipeline that automatically creates toy models to work on that, despise the large amount of variables and complexities, produces good results. From a preliminary analysis of the final results produced by the entire pipeline, if one calculates how many edges are both correctly oriented and signed using Kegg and SignaLink as a reference, out of 200 generated models of about 20 edges each, 75% of the times randomly removed directionality and signs were correctly predicted suggesting that this process is a good starting point for future developments and improvements.